\documentclass{article}[12pt]
\usepackage{epsfig}
\begin{document}
\begin{center}
\Huge Fluctuation of energy in the generalized thermostatistics\\
\vspace{0.5cm}
{\Large F. Q. Potiguar\footnote{e-mail: potiguar@fisica.ufc.br}, U. M. S. Costa\\}
\vspace{0.5cm}
{\Large\em Universidade Federal do Cear\'a, Departamento de F\'\i sica, Campus do Pici, 60455-760, Fortaleza, Cear\'a, Brasil\\}
\end{center}
PACs: 05.20.-y, 05.70.-a\\

\vspace{1.0cm}
\begin{center}
\section*{Abstract}
\end{center}
We calculate the fluctuation of the energy of a system in Tsallis statistics
following the finite heat bath canonical ensemble approach. We obtain this
fluctuation as the second derivative of the logarithm of the partition function
plus an additional term. We also find an explicit expression for the relative
fluctuation as related to the number of degrees of freedom of the bath and the
composite system.\\

\newpage
\section{Introduction}
Tsallis' statistics, or generalized statistics, represents today a very rich field
of research. Since its proposal in 1988 by Tsallis \cite{Tsa88}, many developments
and applications were made in order to give it a physical background. Some examples
are found in studies of L\'evy flights \cite{ZaA95,PrT99}, particle physics
\cite{RWW01,Bec02}, and turbulence \cite{Bec02}. The interpretation relates the
parameter $q$ to the fluctuation of some other important physical property
\cite{Ola01,WiW00}. Hence, many researchers consider it as the right framework
to describe these phenomena.\\
We follow the canonical ensemble with the finite heat bath approach. This approach
was first proposed by Plastino and Plastino \cite{PlasPlas94} and further confirmed by
Almeida \cite{Alm01}, who gave it a strong mathematical background. In this last
reference, the author introduced the equation:
\begin{equation}
\label{finite-bath-condition}
\frac{d}{dE_i}\left(\frac{1}{\beta_i}\right)=q_i-1
\end{equation}
where $\beta_i$ is given by the Boltzmann principle allowing the calculation of
the $q$ parameter of Tsallis' statistics, for any system, isolated or not. We adopt
a convention in the present work: since we are dealing with a canonical ensemble,
we have three systems, the composed, the studied and the heat bath. Then, any
quantities pertaining to these systems are labeled by $0$, $1$ and $2$, respectively.\\
This is a natural consequence of the structure function formalism \cite{Khinchin},
and its connexion to Tsallis' statistical mechanics can be better understood in the
following references, where some issues were treated recently:
\cite{APC02,Alm02,PoC02}.\\
In this work, another issue is tackled. We propose to generalize the calculations
of the energy fluctuation within the canonical ensemble. Since the interpretation
of the parameter $q$ relies on fluctuations, we wish to calculate the fluctuation
of an important quantity in a canonical ensemble, the system's energy $E_1$.
But first, we will revisit the main results in the ordinary picture.\\
In the Boltzmann-Gibbs statistics, particularly the canonical ensemble approach,
we have a general composed system and we wish to study just a small part of it.
The rest of the composed system is called the heat bath, and it is responsible for
maintaining the temperature constant. Since the temperature is a constant, so is
the small part average energy, but the system energy still fluctuates around the
mean value, and the quantities defined previously can give an estimate of these
fluctuations. If they are sufficiently small, we can have great confidence in the
theory results.\\
The dispersion of the energy is given, in the canonical ensemble, by the following
relation, \cite{Reif}:
\begin{equation}
\label{dispersion-BG}
<\Delta E^2>=\frac{\partial^2lnZ}{\partial^2\beta}
\end{equation}
where $Z$ is the partition function and $\beta=\frac{1}{k_BT}$ and
$<\Delta E^2>=<E^2>-<E>^2$. The relative fluctuation reads,
\cite{Khinchin,Landau,Huang}:
\begin{equation}
\label{relative-fluctuation-BG}
\frac{<\Delta E^2>^{\frac{1}{2}}}{<E>}\sim\frac{1}{\sqrt{N}}
\end{equation}
Therefore, we can conclude that if the number of particles of the system is very
great, then its energy will be almost all the time very close to its mean value
$<E>$, hence confirming the validity of statistical mechanics.\\
We will show that, in the generalized picture, the numbers of degrees of freedom
of the heat bath and of the composite system play a more important role in the
measure of these fluctuations.\\
Another important result will be related to the derivative of $lnZ_1$. We will
show that the dispersion of energy is still given by equation (\ref{dispersion-BG}),
but now we have the appearance of other term, proportional to $<E_1>^2$.\\
All these developments will be made in section 2. In section 3, we will illustrate
these ideas with a very simple situation, two ideal gases in equilibrium. Finally,
we will address our conclusions.\\

\section{Calculation of fluctuations}
In Tsallis' picture we have a different energy probability density, which is not
given by an exponential function, but by a power law in the form:
\begin{equation}
\label{Tsallis-distribution}
P(E_1)=\frac{1}{Z_1}\left[1-(q_2-1)\beta^*E_1\right]^{1/(q_2-1)}\Omega_1(E_1)
\end{equation}
where the parameter $\beta^*$ is related to the physical temperature by:
\[
(q_2-1)\beta^*=\frac{q_0-1}{q_0}\frac{1}{k_BT}
\]
and we put a star on it to stress that it is not the Lagrange multiplier
associated with the entropy optimization problem, as stated in \cite{TMP98}.
$Z_1$ is the partition function, and $\Omega_1(E_1)$ is the system's structure
function. In general the structure function is defined as:
$$\Omega_i(E_i)=\frac{q_i}{q_i-1}C_iE_i^{\frac{1}{q_i-1}},$$
where the label $i$ represents the i-system\\
As was already stated, we do not use $q$-average values \cite{TMP98}, we choose
the ordinary one, defined by the following relation:
\begin{equation}
\label{mean-value}
<f(E_1)>=\int f(E_1)P(E_1)dE_1
\end{equation}
where this integral, and all others to come, unless otherwise noted, must be
calculated with the condition on the energy that avoids the distribution $P(E_1)$
to have negative values, the Tsallis' cut-off condition. This condition has a simple
physical interpretation in this approach: it means that the system's energy cannot be
greater than the energy of the whole composite system, that is $E_1\leq E_0$,
\cite{PlasPlas94}.\\
Let us calculate the energy mean-square fluctuation of the energy. First of all,
we know that the temperature, $k_BT$, is proportional to the energy per degree of
freedom of a system \cite{APC02}:
\begin{equation}
\label{temperature}
k_BT=\frac{E_0}{q_0/(q_0-1)}=\frac{<E_1>}{q_1/(q_1-1)}=\frac{<E_2>}{q_2/(q_2-1)}
\end{equation}
This is the well known equipartition theorem \cite{Khinchin}.\\
Hence, the energy second moment is calculated as follows: using
(\ref{Tsallis-distribution}), the relation for $\beta ^*$ and the system structure
function:
\[
<E_1^2>=\frac{\int\left(1-\frac{E_1}{E_0}\right)^{\frac{1}{q_2-1}}
E_1^{\frac{1}{q_1-1}+2}dE_1}{\int\left(1-\frac{E_1}{E_0}\right)^
{\frac{1}{q_2-1}}E_1^{\frac{1}{q_1-1}}dE_1}
\]
These integrals yield beta functions:
\begin{equation}
\label{second-energy-moment}
<E_1^2>=E_0^2\frac{B\left(\frac{q_1}{q_1-1}+2,\frac{q_2}{q_2-1}\right)}
{B\left(\frac{q_1}{q_1-1},\frac{q_2}{q_2-1}\right)}
\end{equation}
With the knowledge of this quantity, we are able to calculate the energy
dispersion. By using the properties of the beta functions and equation
(\ref{mean-value}), we obtain for $<(\Delta E_1)^2>$:
\[
<(\Delta E_1)^2>=\frac{q_0/(q_0-1)}{q_1/(q_1-1)}
\frac{q_1/(q_1-1)+1}{q_0/(q_0-1)+1}<E_1>^2-<E_1>^2
\]
Using the additivity of the degrees of freedom:
\begin{equation}
\label{degrees-freedom}
\frac{q_0}{q_0-1}=\frac{q_1}{q_1-1}+\frac{q_2}{q_2-1}
\end{equation}
we obtain the relative fluctuation of $E_1$:
\begin{equation}
\label{relative-fluctuation}
\frac{<(\Delta E_1)^2>^{\frac{1}{2}}}{<E_1>}=\left[\frac{q_2/(q_2-1)}
{q_0/(q_0-1)+1}\frac{1}{q_1/(q_1-1)}\right]^{\frac{1}{2}}
\end{equation}
Comparing to equation (\ref{relative-fluctuation-BG}) we see that it still
preserves the inverse square root dependence in the number of components of the
studied system, since the quantity $q_1/(q_1-1)$ is proportional to it. But,
now, this constant of proportionality is not trivial. We have a dependence on
the degrees of freedom of the whole system and the bath, represented by the $q_0$
and $q_2$. Although this happens, we still arrive at relation
(\ref{relative-fluctuation-BG}) in the limit $q_2\rightarrow 1$ since, by equation
(\ref{degrees-freedom}), when the above mentioned limit is taken, the limit
$q_0\rightarrow 1$ is implied.\\
There is another way of calculating this relative fluctuation, which is to relate
it to the system's partition function $Z_1$. This is the standard procedure in
Boltzmann-Gibbs (BG) statistics. We start with the definition of the energy second
moment:
\[
<E_1^2>=\frac{\int\left(1-\frac{q_0-1}{q_0}\frac{E_1}{k_BT}\right)^
{\frac{1}{q_2-1}}E_1^2\Omega_1(E_1)dE_1}{\int\left(1-\frac{q_0-1}{q_0}
\frac{E_1}{k_BT}\right)^{\frac{1}{q_2-1}}\Omega_1(E_1)dE_1}
\]
We know that:
\[
<E_1^2>=-\frac{q_2-1}{q_2}\frac{q_0}{q_0-1}\frac{1}{Z_1}\frac{\partial}
{\partial(1/k_BT)}\int\left(1-\frac{q_0-1}{q_0}\frac{E_1}{k_BT}\right)^
{\frac{q_2}{q_2-1}}E_1\Omega_1(E_1)dE_1
\]
This expression can be rewritten as:
\[
<E_1^2>=-\frac{q_2-1}{q_2}\frac{q_0}{q_0-1}\frac{1}{Z_1}\frac{\partial}
{\partial(1/k_BT)}\times
\]
\[\int\left(1-\frac{q_0-1}{q_0}\frac{E_1}{k_BT}\right)E_1\left(1-\frac{q_0-1}{q_0}
\frac{E_1}{k_BT}\right)^{\frac{1}{q_2-1}}\Omega_1(E_1)dE_1
\]
Using equations (\ref{mean-value}) and (\ref{temperature}), we arrive at the
dispersion of the energy:
\begin{equation}
\label{energy-rms}
<(\Delta E_1)^2>=\frac{\partial^2 lnZ_1}{\partial(1/k_BT)^2}-
\frac{q_1/(q_1-1)+1}{q_1/(q_1-1)[q_0/(q_0-1)+1]}<E_1>^2
\end{equation}
We notice the difference between this result and the usual BG one, the last term
on the right hand side of equation (\ref{energy-rms}). When we take the infinite
bath limit, $q_2\rightarrow1$, the total number of degrees of freedom goes to
infinity as well, by relation (\ref{degrees-freedom}) $q_0\rightarrow1$, and this
last term vanishes.\\
It is a simple algebra task to show that this equation leads to relation
(\ref{relative-fluctuation}) for the system's relative fluctuation, assuming the
equipartition of energy.\\
It was shown \cite{PoC02} that the usual mean values have the same relations with
the partition function within this approach and the BG one. All of them are related
in the same fashion to $lnZ_1$ and its derivatives. Here, the dispersion of energy is
still related to the second derivative of $lnZ_1$, but another term comes up,
which is a function of the degrees of freedom of the system and of the composite one.
Why we have such a difference? The answer lies in the fact that this approach and the
BG one have identical energy first moments, equipartition of energy, but this fact
does not repeat in higher moments. This can be shown by a direct calculation of the
$n$th moment of the energy within both approaches. We will start with the Tsallis
formalism.\\
The $n$th moment of the energy is given by:
\begin{equation}
\label{nth-moment}
<E_1^n>=\frac{\int\left(1-\frac{E_1}{E_0}\right)^{\frac{1}{q_2-1}}E_1^n\Omega_1(E_1)dE_1}{\int\left(1-\frac{E_1}{E_0}\right)^{\frac{1}{q_2-1}}\Omega_1(E_1)dE_1}
\end{equation}
It readily gives:
\begin{equation}
\label{nth-moment-Tsallis}
<E_1^n>=E_0^n\frac{B\left(\frac{q_1}{q_1-1}+n,\frac{q_2}{q_2-1}\right)}{B\left(\frac{q_1}{q_1-1},\frac{q_2}{q_2-1}\right)}
\end{equation}
Using the properties of the beta function, we write this expression as:
\[
<E_1^n>=\frac{\left(q_1/(q_1-1)+n-1\right)...\left(q_1/(q_1-1)+1\right)q_1/(q_1-1)}{\left(q_0/(q_0-1)+n-1\right)...\left(q_0/(q_0-1)+1\right)q_0/(q_0-1)}E_0^n
\]
In the BG framework, we define the mean value of a function as:
\[
<f>=\frac{\int e^{-\beta E}f\Omega(E)dE}{\int e^{-\beta E}\Omega(E)dE}
\]
Here, we do not need to worry with the cut-off condition, since the whole energy
is infinite, according to the BG approach. Then, the $n$th moment of the energy
is given by the next equation:
\[
<E^n>=\frac{1}{\beta^n}\frac{\Gamma\left(\frac{q_1}{q_1-1}+n\right)}
{\Gamma\left(\frac{q_1}{q_1-1}\right)}
\]
Once again, using the properties of the gamma function and the value of $\beta$,
we have:
\begin{equation}
\label{nth-moment-BG}
<E^n>=\left(q_1/(q_1-1)+n-1\right)...\left(q_1/(q_1-1)+1\right)q_1/(q_1-1)(k_BT)^n
\end{equation}
Indeed, equations (\ref{nth-moment}) and (\ref{nth-moment-BG}) have equal energy
first moments. In fact, for $n=1$ in (\ref{nth-moment}), we have:
\[
<E_1>=\frac{q_1}{q_1-1}\frac{E_0}{q_0/(q_0-1)}=\frac{q_1}{q_1-1}k_BT
\]
with the help of equation (\ref{temperature}). Also, for $n=1$, equation
(\ref{nth-moment-BG}) yields:
\[
<E_1>=\frac{q_1}{q_1-1}k_BT
\]
We note a close resemblance with expression (\ref{nth-moment-Tsallis}). In fact,
the $n$th power of $k_BT$ is the approximate expression of the product:
\[
(k_BT)^n=\lim_{q_0\rightarrow1}\frac{E_0^n}{\left(q_0/(q_0-1)+n-1\right)...
\left(q_0/(q_0-1)+1\right)q_0/(q_0-1)}
\]
This happens because when we take this limit, the quotient $\frac{q_0-1}{q_0}$
becomes much larger than the greatest value of $n$ in the product,
provided we choose $n$ not to be of the magnitude of the quotient.\\
Now if we allow the number of degrees of freedom of the system to be also great,
much larger than the number $n-1,$ we end up with the final moment in both approaches:
\[
<E_1^n>=<E_1>^n \; ,
\]
recovering the equivalence between the canonical and the microcanonical ensembles.
\section{Application}
To illustrate our results, we take a system which is composed of two ideal gases.
This composed system has $\frac{q_0}{q_0-1}$ degrees of freedom, and each component
gas has $\frac{q_1}{q_1-1}$ and $\frac{q_2}{q_2-1}$ degrees of freedom. Before we
proceed, a word must be said about the treatment of this system in Tsallis' framework.
Within our approach, any system which is in thermal equilibrium with a {\em finite}
heat bath, i. e., a bath where $q_2\neq1$, has a probability density given by
equation (\ref{Tsallis-distribution}). Hence we do not need to have non-trivial
effects, such as long range interactions or fractality for a system to exibith power
law behavior. There is a common belief that these last effects are responsible for
such framework to arise.\\
Our approach relies on another fact, the finiteness of the heat bath. Therefore, the
main point in our framework is that {\em any system in thermal equilibrium with a
heat bath which have a finite number of degrees of freedom, $q_2\neq1$, has an energy
density given by Tsallis power law function}. If we take two finite ideal gases and
make them interact and achieve the equilibrium situation, then we can use Tsallis
statistics to study this problem.\\
We know, from the geometry of the phase space \cite{Khinchin} and equation
(\ref{finite-bath-condition}), that the following relations are valid for an ideal
gas:
\begin{equation}
\frac{q_i}{q_i-1}=\frac{DN_i}{2}
\end{equation}
where $N_i$ is the number of particles in the $i$-th gas and $D$ is the spatial
dimension.\\
Using equation (\ref{relative-fluctuation}), we have:
\begin{equation}
\frac{<(\Delta E_1)^2>^{\frac{1}{2}}}{<E_1>}=\left(\frac{N_2}{\frac{D}{2}N_0+1}
\right)^\frac{1}{2}\frac{1}{\sqrt{N_1}}
\end{equation}
which is still proportional to the inverse square root of the number of particles
in the studied gas. It is important to strees two properties of this relative
fluctuation.\\
The first one regards the infinite bath limit, $N_2\rightarrow\infty$. When we take
this limit, it is implied that the total number of particles oes to infinity as well,
$N_0\rightarrow\infty$. Then, the proportionality constant will be reduced to
$\left(\frac{2}{D}\right)^{\frac{1}{2}}$, which is the constant obtained in the
BG context.\\
The second regards the magnitude of the number of particles. If $N_1$ and $N_2$
are of the same order of magnitude, this relative fluctuation will be proportional
to the inverse of the total number of particles. Therefore, it will only be small if
this total number of particles is large. For example, if we have the numbers: $N_1=2$,
$N_2=4$ and $D=2$, the fluctuation of energy will be given by:
\[
\frac{<(\Delta E_1)^2>^{\frac{1}{2}}}{<E_1>}=\left(\frac{2}{7}\right)^{\frac{1}{2}}
\sim 0.534
\]
This means that when we decrease the number of degrees of freedom of the thermal bath
the energy flutuaction increases. As another exemple if we take a system with $N_1=2000$,
$N_2=4000$ and $D=2$ we obtain:
\[
\frac{<(\Delta E_1)^2>^{\frac{1}{2}}}{<E_1>}\sim 0.018,
\]
and now the energy fluctuation is small.
\section{Conclusions}
We showed that the relative fluctuation of the energy of a system in thermal
equilibrium with a finite heat bath is given by a product of the number of degrees
of freedom of the system, the heat bath, and the whole composite system.\\
We also showed that the dispersion of the energy still have a dependence on the
second derivative of $lnZ_1$, but now a new term arises. This term is proportional
to the square of the mean energy and vanishes properly in the infinite bath limit
$q_2\rightarrow1$.\\
These results show that, within this framework, the role played by the heat bath
is more explicit than in the BG one. In this picture, the bath has the function of
keeping the temperature constant and no other quantities related to it enter into the
equations of the system with which it is in contact. Here, the number of degrees of
freedom of the bath must be known in order to calculate the fluctuations of the
energy. This fluctuation cannot be considered small for any system, only for large
ones when this is an obvious fact.\\
We treated an example here, two finite ideal gases in themal equilibrium. The
relative fluctuation of the system energy depends on the number of particles of
the bath, $N_2$, and of the composed system, $N_0$. This fluctuation reduces
to the result in BG statistics in the infite bath limit, $N_2\rightarrow\infty$.
Also, if the number of particles of both gases are of the same magnitude, the
relative fluctuation is only small if the total number of particles is large.\\

\section{Acknowledgments}
This work was supported by FUNCAP and CNPq. We also acknowledge J.E.Moreira for
a critical reading of the manuscript

\end{document}